\documentstyle{article}

\newcommand{\dar}{Darboux coordinates}
\title{Semiclassical approximation in Batalin-Vilkovisky formalism.}
\author{Albert Schwarz\\
Department of Mathematics, University of California,\\ Davis, CA
95616\\
ASSCHWARZ@UCDAVIS.EDU}

\begin {document}
\maketitle
\smallskip
\begin {abstract}
   The geometry of supermanifolds provided with a $Q$-structure (i.e.
with
 an odd vector field $Q$ satisfying $\{ Q,Q\} =0$), a $P$-structure (odd
symplectic structure ) and  an $S$-structure (volume element) or with
various
combinations of these structures is studied. The results are applied to
the
analysis of the Batalin-Vilkovisky approach to the quantization of
gauge
theories.
In particular the semiclassical approximation in this approach is
expressed in
terms of Reidemeister torsion.
 \end {abstract}
 \vskip .1in
\centerline {\bf {0. Introduction.}}
\vskip .1in
The Batalin-Vilkovisky formalism (BV-formalism)
is based on a notion of an odd Poisson bracket
(antibracket). The odd Poisson bracket of two functions $F,G$ on the
(super)space $R^{n,n}$ with even coordinates $x^1,...,x^n$ and odd
coordinates $\xi_1,...,\xi_n$ can be defined by the formula
\begin {equation}
\{ F,G\} ={\partial F\over \partial x^a}{\partial _lG\over
\partial \xi _a}-{\partial _rF\over \partial \xi _a}
{\partial G\over \partial x^a}
\end {equation}
The transformations of $R^{n,n}$ preserving the bracket (1) are called
odd
symplectic transformations or $P$-transformations. Volume preserving
$P$-transformations are called $SP$-transformations. A manifold $X$
pasted together from domains in $R^{n,n}$ by means of
$P$-transformations is called an odd symplectic manifold or a
$P$-manifold. Replacing $P$-transformations by
$SP$-transformations in this
definition we get a notion of an $SP$-manifold. In a $P$-manifold $X$
we
have a notion of an odd Poisson bracket $\{ F,G\} $ of two functions on
$X$; in an arbitrary coordinate system we can express $\{ F,G\} $ as
$$ \{ F,G\} ={\partial _rF\over \partial z^i}\omega^{ij}(z)
{\partial_lG\over \partial z^j}.$$
The $2$-form $\omega =dz^i\omega_{ij}(z)dz^j$ is closed. (Here the
matrix
$\omega_{ij}$ is inverse to $\omega^{ij}$.)  The formula
$$ \omega (\zeta ,\tilde {\zeta})=\zeta ^i\omega_{ij}(z)\tilde {\zeta}^j$$
determines an odd inner product in the space $T_z(X)$ of tangent
vectors to
$X$ at the point $z\in X$. A Lagrangian submanifold $L$ of $X$ is by
definition
a
$(k,n-k)$-dimensional submanifold of $X$ where the form $\omega$
vanishes
(i.e. $\omega (\zeta,\tilde{\zeta})=0$ for all pairs
$\zeta ,\tilde {\zeta}$ of tangent vectors $\zeta ,\tilde {\zeta}\in
T_z(L)$).
In other words $L$ is a $(k,n-k)$-dimensional isotropic submanifold of
$X$.

 The volume element in an $SP$-manifold $X$ generates a volume
element
$\nu$ in every tangent space $T_z(X)$, i.e. a function $\nu
(e_1,...,e_{2n})$
satisfying $\nu(\tilde{e}_1,...,\tilde{e}_{2n})=\det a\cdot
 \nu (e_1,...,e_{2n})$.

(Here $e_1,...,e_{2n}$ and $\tilde {e}_1,...,\tilde {e}_{2n}$ are bases of
$T_z(X)$ connected by the formula $\tilde {e}_i=a_i^je_j$.). One can
define a
volume element $\lambda$ in a Lagrangian manifold $L\subset X$;
namely if
$e_1,...,e_n$ is a basis of $T_zL$ we take
$$\lambda (e_1,...,e_n)=\mu(e_1,...,e_n,f^1,...,f^n)^{1/2}$$
where $f^1,...,f^n$ satisfy $\omega(e_i,f^j)=\delta_i^j$. We
will assume always that all manifolds under consideration are compact.
Let us define an operator $\Delta$ on an $SP$-manifold $X$ by the
formula
$$\Delta H=div\  K_H$$
where the divergence is determined by the volume element in $X$ and
$K_H^i=\omega^{ij}\partial_lH/\partial z^j$. The Batalin-Vilkovisky
approach
to quantization of gauge theories is based on the consideration of the
integrals having the form
\begin {equation}
 \int\limits _L Hd\lambda
\end {equation}
where $\Delta H=0$ and $L$ denotes the Lagrangian submanifold of
$X$.
It is proved in [1] that (2) does not change by a continuous deformation
of $L$;
more generally (2) does not change when $L$ is replaced by another
Lagrangian
manifold $L^{\prime}$ from the same homology class; see [2]. (Saying
that
two submanifolds $L$ and $L^{\prime}$ of a supermanifold $X$ belong
to the
same homology class we have in mind that their bodies $m(L)$ and
$m(L^{\prime})$ are homologous in the body $m(X)$ of $X$). In
physical
applications the integrand $H$ in (2) has the form $H=\exp (-\hbar
^{-1}S\}$
where $S$ is an extension of the classical action; the choice of
the Lagrangian
manifold $L$ corresponds to the choice of the gauge condition.

  One of our aims is to calculate the asymptotic behavior of $\int_L\exp
(-\hbar^{-1}S)d\lambda$ as $\hbar\rightarrow 0$ (semiclassical
approximation). It is important
to emphasize that the gauge condition (the choice of Lagrangian
submanifold) will not enter directly into the answer. One of
the possible expressions of the answer involves a generalization of
the so called Reidemeister torsion.
The present paper is devoted to the case of finite-dimensional
integrals; however one can develop a similar technique for infinite-
dimensional integrals. This technique is based on the results about
the infinite-
dimensional version of Reidemeister torsion (Ray-Singer torsion)
proved
in [4].  In particular it is useful for the calculation of anomalies in
the BV-approach.

  Our proofs are based on some general results about $SP$-manifolds
and
about Reidemeister torsion. These results are interesting
independently
of applications to the study of semiclassical approximation. One can
consider the corresponding part of the present paper as the next step
in the
analysis of the geometry of BV-quantization after the paper [2].

  The paper is organized as follows. In Section 1 we state the formula
for semiclassical approximation in BV-formalism and give a proof of
this formula. The proof is based on the theorems proved later,
therefore
the reader interested not only in the main ideas, but also in complete
understanding of the proof is recommended to read at first Sections
2-7
and then come back to Section 1. Sections 3-6 are devoted to the
theory of
(super)manifolds provided with  $Q$-structure
(i.e. with an odd vector field $Q$
satisfying $\{ Q,Q\}=0$), $P$-structure (odd symplectic structure) and
 $S$-structure (volume element) or with various combinations of
these structures. The structures studied in these sections can be
considered as a geometric basis of the BV-formalism. For example,
$QP$-structure is connected with master equation and
$QSP$-structure
is related to the quantum version of this equation. In Section 2 we
study
a generalization of Reidemeister torsion. In Section 7 we analyze
the torsion of operators acting on linear $P$-manifolds.

  In the present paper we use the definitions and results of the
paper [2], however one can read this paper independently of [2].

{\bf Notations.} The symbol $R^{p,q}$ denotes $(p,q)$-dimensional
linear
superspace.  The body of a supermanifold $X$ will be denoted by
$m(X)$.
The tangent space to $X$ at the point $x\in X$ is denoted by $T_xX$.
We almost always omit the prefix "super". In particular we write
"manifold"
instead of "supermanifold", "determinant" instead of
"superdeterminant"
(Berezinian). The notations $H_k(X),\  H^k(X)$ are used for homology
and
cohomology groups of $X$ with coefficients in the field of real
numbers. All
functions, mappings etc. are supposed to be smooth.
\vskip .1in
\centerline {\bf 1. Semiclassical approximation.}
\vskip .1in
  Let us consider an $SP$-manifold $X$, a function $S$ on $X$ and a
Lagrangian
submanifold $L\subset X$.

  Let us suppose that $H=\exp(-\hbar^{-1}S)$ satisfies the equation
  $\Delta H=0$ for every $\hbar$; then $S$ obeys the equation $\{
S,S\} =0$
(master equation) and the condition $\Delta S=0$. Let us denote by
$R$ the set
of critical
points of the restriction $\tilde {S}$ of $S$ to the Lagrangian manifold
$L$.
One can
check that the vector field $K^i_S=\omega^{ij}\partial_lS/\partial z^j$
corresponding
to $S$ is an odd vector field satisfying $\{ K_S,K_S\} =0$ and tangent
to $R$
(i.e.
in the terminology of Section 3, $K_S$ determines $Q$-structure both
in $X$ and
$R$.)
The set of zeros of the vector field $K_S$ coincides with the set $Y$ of
all
critical points of $S$. We would like to study the asymptotic behavior of
the
integral of $H$ over $L$ as $\hbar \rightarrow 0$. It is well known that
this
behavior can be expressed in terms of an integral over $R$ (see
Lemma 2).
The integrand of this integral is $K_S$-invariant, hence the integral
over $R$
can be reduced to the integral over a neighborhood of $R\cap Y$
(see[5] or
Lemma 5).
Therefore in principle we can express the
asymptotic behavior of $\int _L\exp (-\hbar ^{-1}S)d\lambda$ as $\hbar
\rightarrow 0$ in terms of the behavior of the function $S$ in the
neighborhood of the set $Y$ of all critical points of $S$. Let us give
the corresponding
expression under some conditions on the dimension of $L$. For every
point $x\in Y$ we define an operator $Q_x$ acting on the tangent
space $T_xY$
as an operator with the matrix $\omega^{ij}S_{jk}$ where $S_{jk}$
denotes
the Hessian of $S$ at the point $x$. It is easy to check that $Q_x^2=0$
and therefore we can consider the homology group $H_x=Z_x/ B_x$
where $Z_x=Ker\  Q_x,\ \  B_x=Im\ Q_x$. It is easy to check that $Z_x
\subset T_xY$; we will suppose that $Y$ is a manifold and
$Z_x=T_xY$. Therefore there exists a natural map of $T_xY$ onto
$H_x$;
we will denote this map by $\pi_x$. The dimensions of $X$ and $Y$
will
be denoted by $(n,n)$ and $(z_1,z_2)$ correspondingly, the dimension
of $B_x$
will be denoted by $(b_1,b_2)$. It is important to emphasize that in the
situations arising in physics the set $Y$ is not necessarily a manifold.

  Let us introduce the notion of a regular critical submanifold of $X$ as
follows. We will say that $C$ is a regular critical submanifold if $C$
is a compact submanifold of $Y$ and for every point $x\in C$ one can
find vectors $e_1,...,e_q,f^1,...,f^q\in T_xY$ in such a way that
$e_1,...,e_q\in T_xC,\ \  \omega (e_i,e_j)=0,\ \  \omega (e_i,f^j)=
\delta_i^j$ and the vectors $\pi_x(e_1),...,\pi_x(e_q),
\pi_x(f^1),...,\pi_x(f^q)$ constitute a basis in $H_x$. It
follows from this definition that for every $x\in C$ one can find
vectors $g_1,...,g_{n-k}\in T_xX$ in such a way that the vectors
$(e_1,...,e_q,f^1,...,f^q,g_1,...,g_{n-q},Q_xg_1,...,Q_xg_{n-q})$
form together a basis in $T_xX$. This remark permits us to define
a volume element in $T_xC$ (and therefore in $C$) by the formula
\begin {equation}
\gamma (e_1,...,e_q)=\mu (e_1,...,e_q,f^1,...,f^q,g_1,...,g_{n-q},
Q_xg_1,...,Q_xg_{n-q})^{1/2}
\end {equation}
Therefore we can talk about the volume of a regular critical manifold
$C$.

{\bf Theorem 1.} The asymptotic behavior as $\hbar \rightarrow 0$ of
the integral of $\exp (-\hbar ^{-1}S)$ over an $(l,n-l)$-dimensional
Lagrangian manifold $L$ can be described as follows:
\begin {equation}
\int\limits _L\exp (-\hbar ^{-1}S)d\lambda\approx \hbar ^{(z_1-z_2)/2}
\int\limits _C\exp(-\hbar^{-1}S)d\gamma
\end {equation}
where $C$ is an appropriate regular critical submanifold of $X$. More
precisely, the homology class $[C]$ of $m(C)$ must be connected with
the homology class $[L]$ of $m(L)$ by the formula
\begin {equation}
 D[C]=i^*D[L]
\end {equation}

   Here $D$ denotes the Poincare duality between homology and
cohomology
groups of a manifold and $i^*$ is a homomorphism of cohomology
groups
induced by the embedding $i:Y\rightarrow X$. In such a way $[L]\in
H_l(m(X)),\ \  D[L]\in H^{n-l}(m(X)),\ \  [C]\in H_{z_1+l-n}(m(Y)),\ \
D[C]\in H^{n-l}(m(Y))$. We suppose that $z_2\geq l\geq b_2,\ n-l\geq
b_1$.

 It is obvious that $S$ is constant on every connected component of
$Y$.
Therefore in the case when $C$ is connected or belongs to a
connected
component of $Y$ the right hand side of (4) is equal to
\begin {equation}
\hbar ^{(z_1-z_2)/2}\exp (-\hbar ^{-1}S_0)\cdot  {\rm volume } (C)
\end {equation}
where $S_0$ denotes the value of $S$ on $C$. In the general case we
get a sum over the components of $C$. (Note that we did not assume
that
$S$
is real.)

 Theorem 1 can be formulated in a more invariant form in the following
way.
First of all we note that the restriction $\tilde {\omega}$ to $Y$
of the $2$-form $\omega$ corresponding to the $P$-structure on $X$
is a degenerate odd closed $2$-form on $Y$. Factorizing $Y$ with
respect to null-vectors of $\tilde {\omega}$ we obtain a manifold
$Y^{\prime}$ provided with a non-degenerate odd closed $2$-form
and therefore with a $P$-structure. (The manifold $Y^{\prime}$ does
not
always exist; see Section 5 for details.) One can identify the
tangent spaces to $Y^{\prime}$ with the homology groups $H_x=
Ker\  Q_x/Im\  Q_x,\ \  x\in Y$. This permits us to introduce
a volume element in $Y^{\prime}$ using the construction of torsion.
(This construction allows us to define a volume element in $H_x$
starting with the volume element in $T_xX$; see Section 2). One
can prove that this volume element together with the $P$-structure
in $Y^{\prime}$ determines an $SP$-structure on $Y^{\prime}$; see
Section 6. The image $C^{\prime}$ of a regular critical submanifold
$C$ of $X$ under the natural projection $Y\rightarrow Y^{\prime}$ is
a Lagrangian submanifold of the $SP$-manifold $Y^{\prime}$ (the
manifold $C^{\prime}$ can have self-intersections). Therefore
we can calculate the volume of $C^{\prime}$ as a volume of
a Lagrangian submanifold of an $SP$-manifold. This permits us to
give an alternative formulation of Theorem 1; namely in the case
when $S$ is constant on $C$ we can use (6) replacing the volume
($C$) by the volume ($C^{\prime}$). (Combining the definition
 of torsion and the definition of a volume element in a Lagrangian
 submanifold we obtain that volume($C$)=volume($C^{\prime}$)).

 The reformulation of the Theorem 1 permits us to use the results
 of [2] to prove this theorem. Namely it follows immediately
 from Theorem 1 of [2] that the left hand side of (4) depends only
  on the homology class of Lagrangian manifold L and the right
  hand side of (4) depends only on the homology class of the
  regular critical submanifold $C$.

 Now we can apply the following lemma.

{\bf Lemma 1.}

  Let us consider an $(n,n)$-dimensional $P$-manifold $X$ and a
function
$S$ on $X$, satisfying $\{ S,S\} =0$. Let us suppose that a Lagrangian
submanifold $L$ of $X$ is in general position with respect to the
manifold
$Y$ of critical points of $S$ (i.e. $L\cap Y$ is a manifold and for
$x\in L\cap Y$ we have $T_x(L\cap Y)=T_xL\cap T_x Y,\ \
T_xL+T_xY=T_xX$).
Then $L\cap Y$ is a regular critical submanifold of $X$.

  To prove this  lemma we note that a non-zero vector $b\in B_x,\ \  x\in
L\cap
Y$
cannot belong to $T_x(L\cap Y)$. Really, $b\in B_x$ is orthogonal to
$Z_x=T_xY$. If $b\in T_xL$, it is orthogonal to $T_xL$ because
$T_xL$
is isotropic. We see that a vector $b\in B_x\cap T_xL$ is orthogonal
to $T_xX=T_xL+T_xY$; hence $b=0$ (because the inner product in
$T_xX$
is non-degenerate). The manifold $L\cap Y$ is isotropic; it follows
from the consideration above that its image $\pi (L\cap Y)$ under the
natural projection $\pi:Y\rightarrow Y^{\prime}$ is an isotropic
manifold of the same dimension (more precisely $\pi$ is an immersion
of $L\cap Y$ into $Y^{\prime}$). Calculating the dimension of $L\cap
Y$
we conclude that $\pi(L\cap Y)$ is a Lagrangian submanifold of
$Y^{\prime}$
and therefore $L\cap Y$ is a regular critical submanifold. (If dim $X=
(n,n)$, dim $Y=$dim $Z_x=(z_1,z_2)$, dim $B_x=(b_1,b_2)$, dim
$L=(l,n-l)$
we have dim $L\cap Y=(z_1+l-n,z_2-l)$, dim
$Y^{\prime}=(z_1-b_1,z_2-b_2)=
(z_1+z_2-n,z_1+z_2-n)$. Taking into account that $(z_1+l-n)+(z_2-l)=
z_1+z_2-n$ we see that $\pi (L\cap Y)$ is a Lagrangian submanifold of
$Y^{\prime}$). In the proof above we considered for simplicity the case
when the $P$-manifold $Y^{\prime}$ exists. However this restriction
is unnecessary. The statement of the lemma is local, hence we always
can work in a neighborhood of $x\in L\cap Y$ where our reasoning is
valid.

To prove Theorem 1 we consider at first the case when the Lagrangian
manifold
$L$
is in general position with respect to $Y$ and all critical points of
the restriction of $S$ to $L$ are critical points of
$S$ on $X$ (i. e. belong
to
$Y$). Applying Lemma 2 we express the asymptotics of (4) as $\hbar
\rightarrow
0$ in terms of an integral over the set $R$ of critical points of the
restriction of $S$ to $L$; in the case at hand $R$ coincides with
the regular
critical submanifold $C=L\cap Y$. The integrand can be represented
as a
partition function of a quadratic functional $S_x$ (of the Hessian of
$S$
at the
point $x\in C$); see Lemma 2. This partition function can be expressed
in terms
of torsion; see Lemma 7$^{\prime \prime}$. Using this fact we obtain
the
statement of Theorem 1 (or more precisely, the equivalent statement
formulated
above) in the case at hand. Essentially the same arguments combined
with Lemma 5
can be applied if we replace the condition $R=L\cap Y$ with the
weaker
condition
that for every $x\in C=L\cap Y$ the subspace $N_x$ of zero modes of
the restriction
of $S_x$ to $T_xL$ coincides with $T_xC=T_xL\cap T_xY$. Moreover,
if this
condition is violated on a subset of $C$ having zero measure we still
can
arrive at the same conclusion.

 If $L$ is an arbitrary Lagrangian submanifold of $X$ and $z_2\geq l$
then in
every neighborhood of $L$
we can find a Lagrangian submanifold $\tilde {L}$ in general position
with
respect to $Y$. It follows from Lemma 9 that in the case when $l\geq
b_2,\
n-l\geq b_1$ we can choose $\tilde {L}$ in such a way that the
arguments above
give a proof of (4) with $L$ replaced by $\tilde {L}$ . This proves (4) for
arbitrary $L$, because neither side of (4) changes when we replace
$L$ by
$\tilde {L}$.

 Let us formulate in conclusion a topological theorem leading to
equation
(5). Let us consider an $m$-dimensional manifold $M$ and an
$n$-dimensional
submanifold $N$. Let us suppose that a $k$-dimensional submanifold
$L\subset M$
is in general position with respect to $N$. Then the intersection $L\cap
N$ is
a manifold. We will suppose that $M,\  N$ and $L$ are oriented
compact
manifolds. Then one can consider homology classes $[L]\in H_k(M),\
[L\cap
N]\in H_{k+n-m}(N)$ of $L$ and $L\cap N$ and their Poincare duals
$D[L\cap
N]\in H^{m-k}(N)$ and $D[L]\in H^{m-k}(M)$. It is well known that by
an appropriate choice of orientation of $L\cap N$ we have
 $$D[L]=j^*D[L\cap N]$$
 where $j^*$ denotes the homomorphism of cohomology groups
corresponding to the
embedding $j:\  N\rightarrow M$ (see for instance [6]). To get (5) we
have to
apply this formula to the embedding of the body of $Y$ into the body of
$X$
induced by the embedding $i:\  Y\rightarrow X$. It is important to
emphasize
that the homology classes of regular critical submanifolds arising in the
right
hand side of (4) can be characterized in purely topological terms and
the
$SP$-structure on $X$ does not enter the answer.
  \vskip .1in
\centerline {\bf 2. Torsion. Definition and main properties.}
\vskip .1in
Let $E$ denote a (finite-dimensional) linear superspace over $R$.
A linear
measure $\lambda$ on $E$ is by definition an even function
$\lambda(e)$ of the basis $e$ in $E$ satisfying the condition
$\lambda(Ae)=\det A\cdot\lambda(e)$. (Here $Ae$ denotes the basis
in
$E$ obtained from $e$ by means of the linear transformation $A$. In
other
words the bases $e=(e_1,...,e_k)$ and $\tilde {e}=Ae=
(\tilde {e}_1,...,\tilde {e}_k)$ are connected by the formula
$\tilde {e}_i=A_i^ke_k.$) The one-dimensional space of linear
measures on $E$ will be denoted by $\Lambda(E)$ or simply by
$\Lambda E$. We say that the measure $\lambda$ specifies a volume
element
in $E$ if $\lambda(e)$ is invertible. (Recall that every element
$\lambda$ of a Grassmann algebra can be represented as a sum of its
number part $m(\lambda)$ and its nilpotent part $n(\lambda)$. An
element $\lambda$ is invertible if $m(\lambda)\not= 0$). Note that
every
basis $f$ in $E$ determines a volume element $\lambda_f$ in $E$ by
means of the condition $\lambda_f(f)=1$. It is easy to check that there
exist canonical isomorphisms
\begin {equation}
\Lambda(E^*)=\Lambda(\Pi E)=\Lambda(E)^*,     \label {eq:(1)}
\end {equation}
\begin {equation}
\Lambda(E)=\Lambda(E^{\prime})\otimes\Lambda(E/E^{\prime}).
\label {eq:(2)}
\end {equation}
Here $E^*$ denotes the space dual to $E,\ \Pi E$ is obtained from $E$
by
means of parity reversion and $E^{\prime}$ is a linear subspace of
$E$.
Let us denote by $Q$ a parity reversing linear operator in $E$
satisfying
$Q^2=0$. As usual we can define the subspace of cycles $Z$ as the
kernel
of $Q$ and the subspace of boundaries $B$ as the image of $Q$.
The homology
group $H=H(E,Q)$ can be defined as the linear superspace $Z/B$.
(One
has
to impose some conditions on $Q$ to guarantee that $Z$ ,$B$ and
$H$
can be
considered as superspaces; we always assume that these conditions
are
satisfied.) It is easy to check that there exists a canonical isomorphism
\begin {equation}
\Lambda(E)=\Lambda(H).     \label {eq:(3)}
\end {equation}
This isomorphism will be denoted by Tor, because it can be
considered as
a generalization of Reidemeister torsion. To construct this
isomorphism
we note that it follows from (\ref {eq:(1)}), (\ref {eq:(2)})
and the relations
\begin {equation}
H=Z/B,\ \ \Pi B=E/Z
\end {equation}
that
\begin {equation}
\Lambda(Z)=\Lambda(H)\otimes \Lambda(B),\\
\Lambda(E)=\Lambda(Z)\otimes \Lambda(\Pi B)=\Lambda(Z)\otimes
\Lambda(B)^*.
                  \label {eq:(5)}
\end {equation}
The isomorphism (\ref {eq:(3)}) is an immediate consequence of (\ref
{eq:(5)}).

  Usually the torsion is considered in the case when $H=0$ and $E$ is
provided with a linear measure $\lambda$ from the very beginning.
Then
$\Lambda(H)$ can be identified with $R^{1,0}$ and the torsion can be
interpreted as an even Grassmann number (the image of $\lambda$
by the
isomorphism (\ref {eq:(3)})).

  One can give a more explicit description of torsion. Let us fix vectors
$e_1,...,e_k\in E$ in such a way that $b_1=Qe_1,...,b_k=Qe_k$
constitute
a basis of $B$. If $H=0$ the vectors $e_1,...,e_k,b_1,...,b_k$ form a
basis of $E$ and the torsion can be defined as
\begin {equation}
\lambda(e_1,...,e_k,b_1,...,b_k)   \label {eq:(6)}
\end {equation}
where $\lambda$ is a measure in $E$. In general case we add
elements
$h_1,...,h_s\in Z$ to $e_1,...,e_k,b_1,...,b_k$ to get a basis of $E$.
The images  $\tilde {h}_1,...\tilde {h}_s$ of $h_1,...,h_s$ by the natural
map of $Z$ onto $H$ constitute a basis of $H$. We can define a
measure
$\tilde {\lambda}$ in $H$ corresponding to $\lambda$ by the formula
\begin {equation}
\tilde{\lambda}(\tilde{h}_1,...,\tilde{h}_s)=
\lambda(e_1,...,e_k,b_1,...b_k,h_1,...h_s).
\end {equation}

   For every polyhedron $X$ we can consider the space $E$ of all cell
chains (or cochains) over $R$ in $X$ (the parity is determined by the
dimension of a chain). The boundary (or coboundary) operator
changes the
dimension by one, therefore it is parity reversing and can play the role
of the operator $Q$. The standard basis consisting of cells determines
an
element $\lambda$ of $\Lambda(E)$. Therefore the isomorphism (\ref
{eq:(3)})
determines an element of $\Lambda(H)$ (a number in the acyclic
case). This
construction gives the Reidemeister torsion of $X$. (One can consider
also Reidemeister torsion in a little bit more general situation when
$E$
consists of chains with local coefficients. The local coefficient system is
determined by a representation of $\pi_1(X)$.)

    It is easy to check some simple properties of torsion. Let us consider
 a linear subspace $E^{\prime}\subset E$ invariant with respect to an
 operator $Q$ acting on $E$. Then the operator $Q$ generates an
operator
 $\bar {Q}$ acting on the coset space $F=E/E^{\prime}$. If $Q^2=0$
we
can
 define $H(E^{\prime},Q)$, $H(E,Q)$, and  $H(F,\bar{Q})$. These
homology
 groups are
connected
 by an exact triangle. It follows from this fact that
\begin {equation}
\Lambda H(E,Q)=\Lambda H(E^{\prime},Q)\otimes\Lambda
H(F,\bar{Q})
          \label   {eq:(8)}
\end {equation}
and that
$$H(E,Q)=H(F,\bar{Q})$$
in the case when $H(E^{\prime},Q)=0$. If $Q$ is parity reversing and
the
volume elements in $E,E^{\prime}$ and $F$ are fixed we can define
torsions
of $E,E^{\prime}$ and $F$ as elements of $\Lambda H(E,Q)$,
$\Lambda
H(E^{\prime},Q)$ and $\Lambda H(F,\bar{Q})$ correspondingly. We
will assume
that the volume elements in $E,E^{\prime}$ and $F$ are compatible,
i.e.
the volume element in $E^{\prime}$ and $F$ generate the volume
element in
$E$ by means of (\ref {eq:(2)}). It is easy to prove that in this case

\begin {equation}
Tor(E,Q)=\pm Tor(E^{\prime},Q)\otimes Tor(F,\bar{Q})\label {eq:(9)}
\end {equation}
(In the case when $H(E^{\prime},Q)=0$ and therefore the groups
$H(E,Q)$ and
$H(E,\bar{Q})$ coincide one can replace $\otimes$ by common
multiplication.
In the general case we use the identification (\ref {eq:(8)}).)

  The notion of torsion is closely related to the notion of a partition
function of a degenerate quadratic functional. Let us suppose that $S$
is a quadratic linear functional on a linear superspace $E$ provided
with a volume element $\lambda$. In the case when the quadratic form
$S$ is non-degenerate the partition function $Z_S$ can be defined as
the integral
\begin {equation}
Z_S=\int \limits _E e^{-S}d\lambda=(2\pi)^{d(E)/2}(\det \hat{S})^{-1/2}
\end{equation}
where $d(E)$ denotes the difference between even and odd
dimensions of $E$
and $\hat{S} $ stands for the matrix of quadratic form $S$ in the
basis $e_1,...,e_n$ satisfying $\lambda(e_1,...,e_n)=1$. If $S$
is degenerate we define $Z_S$ as a volume element on
the space $N$ of zero modes of $\hat {S}$. Namely, if $f_1,...,f_k$
is a basis of $N$ we define
\begin {equation}
Z_S(f_1,...,f_k)=Z_{E^{\prime}}
\end{equation}
Here $Z_{E^{\prime}}$ stands for the partition function of $S$
restricted
to the subspace $E^{\prime}\subset E$ satisfying $E^{\prime}+N=E$.
The
measure in $E^{\prime}$ is given by the formula $\lambda^{\prime}
(e_1,...,e_{n-k})=\lambda(e_1,...,e_{n-k},f_1,...,f_k)$. An explicit
formula for $Z_S$ can be written as follows:
\begin {equation}
Z_S(f_1,...,f_k)=(2\pi)^{d(E)-d(N)}(\det \sigma)^{-1/2}
\end {equation}
where $\sigma$ denotes the matrix $\sigma_{ij}=S(e_i,e_j)\ \ 1\leq
i,j\leq
n-k$,
and the vectors $e_1,...,e_{n-k}$ are chosen in such a way that
$\lambda(e_1,...,e_{n-k},f_1,...,f_k)=1$. (Here $S(x,y)$ stands for the
bilinear form corresponding to the quadratic form $S(x)$.)

   Note that the notion of a partition function of a quadratic functional
can
 be used to describe the asymptotic behavior of an integral
 $\int _X\exp (-\hbar ^{-1}{\cal S})d\nu$ as $\hbar \rightarrow 0$. Let
 us suppose that a function ${\cal S}$ is defined on a manifold $X$
 provided with a volume element (on an $S$-manifold). The set of all
 critical points of ${\cal S}$ will be denoted by $R$; for the sake of
 simplicity we assume that $R$ is a compact connected manifold. The
 Hessian of ${\cal S}$ at the point $x\in R$ can be considered as a
 quadratic form ${\cal S}_x$ defined on the tangent space $T_xX$. It
 is clear that the elements of $T_xR$ are zero modes of ${\cal S}_x$;
 we suppose that the space $N_x$ of zero modes of ${\cal S}_x$
 coincides with $T_xR$ (i.e. $R$ is a non-degenerate critical manifold).
 Then the partition function $Z_{{\cal S}_x}$ of a degenerate quadratic
 functional ${\cal S}_x$ can be considered as a volume element on
 $N_x=T_xR$. In such a way we constructed a volume element in $R$;
 we can calculate therefore the volume of $R$.

  {\bf Lemma 2.} The asymptotic behavior as $\hbar \rightarrow 0$ of
the
integral of $\exp (-\hbar^{-1}{\cal S})$ over $X$ can be described
by the formula
$$\int \limits _X\exp (-\hbar ^{-1}{\cal S})d\nu\approx\hbar
^{(d(Y)-d(X))/2}
\cdot \exp(-\hbar^{-1}{\cal S}_0){\rm volume}(R)$$
Here ${\cal S}_0$ denotes the value of ${\cal S}$ on $R$.
(We assumed that $R$ is connected; therefore ${\cal S}$ is
constant on $R$.)

   Let us suppose now that the quadratic form $S$ is invariant with
respect
 to the parity reversing operator $Q$ (this means that $S(z+Qw)=S(z)$
or,
 equivalently, $\hat{S}Q=0$). The space $N$ of zero modes of $S$ is
 invariant with respect to $Q$; we will assume that $Q^2=0$ on $N$.
We can
 consider the homology group $H=H(N,Q)$ of $N$ with respect to $Q$.
We will
 define the partition function $Z_{S,Q}$ as a torsion of $Q$ in $N$ with
 respect to the measure $Z_S$ in $N$. The partition function
$Z_{S,Q}$ is
 a number if $H=0$ and a measure on $H$ (an element of
$\Lambda(H)$) in
the general case. Note that the definition of $Z_{S,Q}$ given above
can
be applied also to the case when the parity reversing operator $Q$
acts only
on $N$ and satisfies $Q^2=0$ there. The definition above is prompted
by
the generalization of the Fadeev-Popov trick introduced in [4]. Let us
suppose
that the quadratic form $S$ on $E_0$ is invariant with respect to a
linear
operator $T_0$ acting from $E_1$ into $E_0$ (i.e. $S(x+T_0y)=S(x)$
for
every $y\in E_1$, or equivalently $Im\ T_0\subset N=Ker\ \hat{S}$).
We
will assume that $N=Im \ T_0$ (i.e. the degeneracy of $S$ is due
entirely
to the symmetry $T_0$) and that there exists a resolution of $T_0$, i.e.
a sequence of spaces $E_0,E_1,...,E_n$ provided with volume
elements and linear maps $T_i:E_{i+1}\rightarrow E_i$ obeying the
condition $Ker\ T_{i-1}=Im\ T_i$. Let us construct a superspace $E$
taking the direct sum $\sum E_{2k}$ as the even part of $E$ and the
direct sum $\sum E_{2k+1}$ as the odd part of $E$. Then the
operators
$T_i$ determine a parity reversing operator $Q$ on $E$ and the form
$S$
determines a form on $E$ that will be denoted by the same letter (by
definition $Q=0$ on $E_0$ and $S=0$ on $E_i,\ i>o$). It follows from
our assumptions that $Q^2=0$ and that the homology group of
$Q$ on the space of zero modes of $S$ is trivial; therefore we can
consider the number $Z_{S,Q}$. It is easy to check that $Z_{S,Q}$
coincides
with the partition function of $S$ with respect to the resolution
$(E_i,T_i)$ as it is defined in [4]. (To be more precise the definition
of [4] is given for the infinite-dimensional elliptic case,
but one can give an
analogous definition in the finite-dimensional case at hand.)
 \vskip .1in
\centerline {\bf 3. $Q$-manifolds.}
\vskip .1in
 Let us consider a supermanifold $X$ provided with an odd vector field
$Q$
 satisfying $\{ Q,Q\} =0$ ( in other words, the corresponding
first order differential operator $\hat{Q}$ obeys
$\hat{Q}^2=0$). We will say
that such a vector field specifies a $Q$-structure on $X$ or that $X$
is a $Q$-manifold. Let us denote the set of all points $x\in X$ satisfying
$Qx=0$ by $Y$. The vector field $Q$ induces a linear map $Q_x$ of
the tangent
space $T_xX$ to $X$ at the point $x\in Y$ into itself. The matrix of
$Q_x$
in local coordinates $(z^1,...,z^n)$ is $\partial Q^a/\partial z^b$.
The map $Q_x$ is parity reversing and obeys
$Q_x^2=0$, therefore we can consider the homology group
$H_x=H(T_x,Q_x)$
and the torsion of $Q_x$. We will suppose that for every $x\in Y$
the kernel $Z_x$ and the image $B_x$ of the operator $Q_x$ are
linear
superspaces of the same dimension and that $Y$ is a supermanifold.
Then the tangent space $T_xY$ to $Y$ at the point $x\in Y$ can
be identified with $Z_x$. The simplest example of a vector field $Q$
satisfying the conditions above is an odd field with constant
coefficients
in $R^{m,n}$ (for example a field with $\hat
{Q}={\partial\over\partial\xi^a}$
where $\xi^a$ denotes one of the odd coordinates in $R^{m,n}$). In
this case
$Y$ is empty. It follows from the superanalog of Frobenius' theorem
that
conversely in the case when $Y$ is empty one can find a
coordinate system
in the $Q$-manifold $X$ such that $\hat {Q}=\partial /\partial \xi^a$. A
more
interesting
example is given by the formula
\begin {equation}
\hat{Q}=\eta^a{\partial\over\partial \xi^a}  \label {eq:(201)}
\end {equation}
where the coordinates $\xi^a$ and $\eta^a$ have opposite parity. It is
clear
that the formula (\ref {eq:(201)}) can be used to determine
a $Q$-structure on
the  supermanifold $TN$ of tangent vectors to a supermanifold $N$
(locally a
point of $TN$ is described by coordinates $\xi^a $ in $N$ and
coordinates
$\eta^a $ of a tangent vector; $\xi^a$ and $\eta^a$ have opposite
parity). This
$Q$-structure will be called the standard $Q$-structure on $TN$.
Differential
and pseudodifferential forms on $N$ can be considered as functions
on $TN$. (By
definition a differential form is a polynomial function with respect to
$\eta^a$
and  a pseudodifferential  form must vanish for $\eta^a$ tending to
infinity
if
$\eta^a$ is even. If all $\xi^a$ are even, all $\eta^a$ are odd, these two
definitions coincide.) The operator (\ref {eq:(201)}) coincides with the
differential of a differential (or pseudodifferential) form. The set $Y$
can
be
identified with $N\subset TN$ in the case at hand. It is easy to check
that
$H(T_x,Q_x)=0$ at every point $x\in N$.

  One can check that the form (\ref {eq:(201)}) of $\hat {Q}$ is general
in
some sense. In particular, one can prove the following

 {\bf Theorem 2.} If $X$ is a $Q$-manifold, $Y=\{ x\in X|Qx=0\}$, then
for
every $x\in Y$ one can find coordinates $(\xi^a,\eta^a,\zeta^{\alpha})$
in a
neighborhood $U$ of $x$ in such a way that $Y\cap U$ is singled out
by the
equations $\eta^a=0$ and $\hat {Q}$ has the form (\ref {eq:(201)}) in
$Y\cap
U$.

  Let us say that a vector field $b$ in $Y$ belongs to the set ${\cal B}$
if
for every $x\in Y$ we have $b(x)\in B_x=Im Q_x$. One can check
that for the
fields $b\in {\cal B},\ b^{\prime}\in {\cal B}$ their (graded) commutator
$[b,b^{\prime}\}\in \cal{B}$. The proof is based on the fact that the
restriction to $Y$ of the field $[Q,A\}$ belongs to ${\cal B}$ for every
vector
field $A$ in $X$ and that locally every field $b\in {\cal B}$ can be
represented as a restriction of a field $[Q,A\}$. This follows
immediately from
the remark that in local coordinates $z^a$ the field $[Q,A\}$ has the
form
$(\partial Q^a/\partial z^b)\cdot A^b$ at the point $x\in Y$. Using the
representations $b=[Q,A\},\  b^{\prime}=[Q,A^{\prime}\}$ and the
Jacoby
identity we obtain
  $$[b,b^{\prime}\}=\pm [Q,[A,b^{\prime}\}\} .$$
 and therefore $[b,b^{\prime}\} \in {\cal B}$. This assertion permits us to
use
the generalization of Frobenius' theorem to the case of a
supermanifold.
Utilizing
this theorem we get a local coordinate system $(u^1,...,u^m)$ in
$Y$ in
so that in the neighborhood of a point $x\in Y$ the field
$b=(b^1,...,b^m)\in
{\cal B}$ if and only if $b^{k+1}=...=b^m=0$. (In other words the
subspaces
$B_x\subset T_xY$ determine a foliation of $Y$.) We can extend this
coordinate
system to a local coordinate system $(u^1,...,u^m,v^1,...,v^s)$ in $X$
in such
a way that $Y$ is singled out by equations $v^1=...=v^s=0$. It is easy
to check
that the operator $\hat {Q}$ in this coordinate system has the form
 $$\hat{Q}=\sum_{a=1}^k\sum_{b=1}^sQ_b^a(u^1,...,u^m)v^b{\partial\o
ver \partial
u^a}+...$$
 where we omitted higher order terms with respect to $v^1,...v^s$. One
can
verify that the matrix $Q_b^a$ must be non-degenerate (and therefore
$s=k$).
  This follows from the remark that the kernel of the matrix $Q_y$ at
the point
$y\in Y$ coincides with $Z_y=TY(y)$. Therefore we can introduce new
coordinates
  $$\xi^a=u^a,\ 1\leq a\leq k,\  \zeta ^{\alpha}=u^{\alpha +k},\  1\leq
\alpha
\leq m-k,\\ \eta^a=Q_b^a(u)v^b,\ 1\leq a\leq k.$$
  We have
 \begin {equation}
 \hat{Q} =\eta^a{\partial\over\partial\xi^a}+({\rm higher\  order\  terms\
with\   respect\   to }\  \eta)     \label {eq:(202)}
 \end {equation}
 in these coordinates. To eliminate the higher order terms in (\ref
{eq:(202)})
we will use the following lemma.

 {\bf Lemma 3. } Let us consider a domain $U$ with coordinates
$(\xi^a,\eta^a,\zeta^{\alpha})$ and an odd vector field  $A$  in $U$
satisfying
$\{Q_0,A\}=0$ and vanishing for $\eta^a=0$. Then $A$ can be
represented as
$A=[Q_0,B]$ where $B$ is an even vector field in $U$ such that
$B=0$ for
$\eta^a=0$.

  Here we used the notation $Q_0$ for an odd vector field
corresponding to the
operator $\hat {Q}_0=\eta^a\partial/\partial\xi^a$. This vector field
determines a $Q$-structure on $U$ and the lemma shows that an
infinitesimal
deformation of this $Q$-structure is trivial (i.e. it leads to an equivalent
$Q$-structure) if the field $Q$ in the deformed $Q$-structure has the
same
zeros as $Q_0$. To prove the lemma we introduce notations
  \begin {equation}
 \hat {A}=A^a{\partial\over\partial
\xi^a}+\tilde{A}^a{\partial\over\partial\eta^a}+
A^{\alpha}{\partial\over\partial\zeta^{\alpha}},       \label {eq:(203)}
 \end {equation}
  \begin {equation}
\hat {B}=B^a{\partial\over\partial
\xi^a}+\tilde{B}^a{\partial\over\partial\eta^a}+
B^{\alpha}{\partial\over\partial\zeta^{\alpha}}   \label {eq:(204)}
 \end {equation}
 and note that the equations $\{ Q_0,A\}=0,\  A=[Q_0,B]$ can be
written as
  \begin {equation}
 \hat{Q}_0\tilde {A}^a=0,\  \hat{Q}_0A^{\alpha}=0,\
\hat{Q}_0A^a-\tilde{A}^a=0,   \label {eq:(205)}
 \end {equation}
  \begin {equation}
 \tilde{A}^a=\hat{Q}_0\tilde{B}^a,\  A^{\alpha}=\hat{Q}_0B^{\alpha},\
A^a=\hat{Q}_0B^a+\tilde{B}^a.     \label {eq:(206)}
 \end {equation}
 We see that one can take $\tilde{B}^a=A^a,\  B^a=0$. To find
$B^{\alpha}$ we
take into account that $\hat{Q}_0$ can be consider as an exterior
differential and
use the Poincare lemma.

  The lemma can be used to eliminate the higher order terms with
respect
to
$\eta^a$ in (\ref {eq:(202)}) by means of an appropriate change of
coordinates in
a neighborhood of $Y$. One can give a heuristic proof of this
statement
eliminating the second order terms, then the third order terms and so
on. To
get a rigorous proof one can write an equation for the change of
coordinates we
are interested in and prove the existence of a solution to this equation
using
the implicit function theorem and Lemma 3. (This Lemma gives us the
information
about the solution to the equation obtained by means of linearization of
the
equation at hand.)

   We proved Theorem 2. The same arguments can be applied to
prove:
   {\bf Theorem 2$^{\prime}$.} Let us suppose that $X$ is a manifold
provided
with a $Q$-structure, $Y=\{x\in X|Qx=0\}$ and $H_x=H(T_x,Q_x)=0$
for $x\in Y$.
Then in a neighborhood of $Y$ this $Q$-structure is equivalent to the
standard
$Q$-structure on $TY$.
 \vskip .1in
   \centerline {\bf 4. $S$-manifolds and $QS$-manifolds.}
\vskip .1in
   To determine a volume element in a manifold $X$ we have to
choose volume
elements in all tangent spaces $T_xX$,  $x\in X$. (We suppose that
these volume
elements depend smoothly on $x\in X$.) We will say that a  volume
element in
the manifold $X$ specifies an $S$-structure on $X$. In local
coordinates
$z^1,...,z^n$ a  volume element is determined by an even invertible
function
$\rho(z)$ (density). By a change of local coordinates the density
gains a
factor equal to a (super)Jacobian.

   The $S$-structure on $X$ determines an operator $div$ acting from
the space
of vector fields in $X$ into the space of functions on $X$. In local
coordinates
   $$div A=\rho ^{-1}{\partial_r\over\partial z^a}(\rho A^a)={\partial_r
A^a\over\partial z^a}+{\partial _r\ln\rho\over\partial z^a}A^a$$
   If the manifold $X$ is provided with a $Q$-structure and an
$S$-structure
simultaneously and the $S$-structure is $Q$-invariant (i.e. $div Q=0$)
we will
talk about a $QS$-structure on $X$.

   As we mentioned above, if $X$ is provided with  $Q$-structure the
subspaces
$B_x\subset T_xY$ determine a foliation of $Y=\{x\in X|Qx=0\}$. Let us
suppose
that this foliation is a fibration of $Y$; the base of this fibration will be
denoted by $R$. The set $C(R)$ of all functions on $R$ can be
identified with
the set of  functions  $\varphi$ on $Y$ satisfying $\hat{b}\varphi =0$ for
all
  vector fields satisfying $b(x)\in B_x,\ x\in Y$ (i.e. for $b\in {\cal B}$).
The set $Vect(R)$ of all vector fields on $R$ can be identified with the
set of
 vector fields $A$ on $Y$ satisfying $[b,A\}\in\cal{B}$ for every $b\in
\cal{B}$. If $x\in Y$ the natural projection $\pi$ of $Y$ onto $R$
induces a
linear map $\pi_*^x$ of the tangent space $T_xY=Z_x$ onto the
tangent
space
$T_{\pi(x)}R$ and the kernel of $\pi_*^x$ coincides with $B_x$. This
means
that exists a natural one-to-one linear map from the homology group
$H(T_x,Q_x)=Z_x/B_x$ onto the  tangent space $T_{\pi(x)}R$. Recall
that there
exists a one-to-one correspondence $\  Tor\  $ between volume
elements in
$T_x=T_x(X)$ and volume elements in $H(T_x,Q_x)$. If $X$ is a
$QS$-manifold
this correspondence gives us a volume element in $T_rR$ for every
$r\in R$ by
means of the identification $T_rR=H(T_x,Q_x)$ where $\pi(x)=r$. We
will
prove that
the volume element in $T_rR$ does not depend on the choice of $x$
and therefore
that $R$ has a natural $S$-structure. It is sufficient to check that the
identifications $T_rR=H(T_x,Q_x)$ and
$T_rR=H(T_{\tilde{x}},Q_{\tilde{x}})$
generate the same volume element in $T_rR$ in the case when
$\tilde{x}$ belongs
to w a neighborhood $U$ of $x\in Y$ such that we can apply Theorem
1.
The
assumption $div\hat{Q}=0$ leads to the equation $\hat{Q}_0\rho =0$ in
the
coordinates $(\xi ,\eta ,\zeta )$. (Here $\hat{Q}_0=\eta^a\partial/\partial
\xi^a$.) Writing the density $\rho$ in the form $\rho=\rho_0(\xi
,\zeta)+\tilde{\rho}(\xi ,\eta ,\zeta)$ where $\tilde{\rho}(\xi ,\eta
,\zeta)=0$ for $\eta=0$ we obtain from $\hat{Q}_0\rho =0$ that
$\rho_0(\xi
,\zeta )$ does not depend on $\xi$. Independence of the volume
element in
$T_rR$ on the choice of $x$ follows immediately from this fact.

   Let us describe the set $C(R)$ of functions on $R$, the set
$Vect(R)$ of
vector fields  on $R$ and the operator $div:\  Vect(R)\rightarrow C(R)$
corresponding to the natural $S$-structure on $R$ in terms of
functions
and
vector fields  on $X$. The operator $\hat{Q}$ can be considered as a
differential acting on the space of functions on $X$ and therefore we
can
consider the homology group ${\cal H}^0(X)={\cal Z}^0(X)/{\cal
B}^0(X)$ where
${\cal Z}^0(X)$ consists of $Q$-closed  functions (i.e. functions
satisfying
$\hat{Q}f=0$) and ${\cal B}^0(X)$ consists of $Q$-exact functions
(functions of
the form $\hat {Q}\varphi$). Analogously, the (super)commutator with
$\hat{Q}$
can be considered as a differential in the space of vector fields on $X$
and we
can consider corresponding homology group ${\cal H}^1(X)={\cal
Z}^1(X)/{\cal
B}^1(X)$. (Here ${\cal Z}^1(X)$ consists of vector fields  $A$ satisfying
$[Q,A\} =0$ and ${\cal B}^1(X)$ consists of vector fields  of the form
$[Q,A\})$. A function $f\in {\cal Z}^0(X)$ restricted to $Y$ can be
considered
as a function on $R$ because $\hat {b}f=0$ for every vector field $b\in
{\cal
B}$. (Locally one can represent $b\in {\cal B}$ as a restriction of
$[Q,a\}$
and therefore $bf=\hat{Q}\hat{a}f$ vanishes on $Y$.)  Functions
belonging to
${\cal B}^0(X)$ vanish on $Y$ therefore the construction above
determines a
linear map from ${\cal H}^0(X)$ into $C(R)$; we will denote this map
by
$\rho_0$. Similarly the restriction of a  vector field $A\in {\cal Z}^1(X)$
to
$Y$ determines a vector field on $R$ and the restriction of a  vector
field
$A\in {\cal B}^1(X)$ vanishes on $R$. Hence we have a linear map
from ${\cal
H}^1(X)$ into $Vect(R)$; this map will be denoted by $\rho_1$. It is
easy to
check that for every point $x\in Y$ there exists a neighborhood
$U$ of $x$
in $X$ such that the maps $\rho_0$ and $\rho_1$ determine
isomorphisms
${\cal
H}^0(U)=C(R_U)$ and ${\cal H}^1(U)=Vect(R_U)$. (Note that the
neighborhood $U$
 always can be chosen in such a way that the foliation of $Y\cap U$ is
a
fibration; the base of this fibration is denoted by $R_U$.) This fact
follows
immediately from Theorem 1 and the Poincare lemma. For example to
prove that
$\rho_1$ is an isomorphism we conclude from (\ref {eq:(205)}), (\ref
{eq:(206)}) that every  vector field belonging to ${\cal Z}^1(U)$ is
homologous
to a  vector field (\ref {eq:(203)}) with $A^a=0,\  \tilde {A}^a=0,\
Q_0A^{\alpha}=0$ and that such a  vector field belongs to  ${\cal
B}^1(U)$  iff
$A^{\alpha}$ can be represented in the form $A^{\alpha}=\hat
{Q}_0B^{\alpha}$.
This remark permits us to identify  ${\cal H}^1(U)$ with the space of
functions
$A^{\alpha}$ depending only on $\zeta$ (i.e. with $Vect\  (R)$).

   It follows from $div\hat {Q}=0$ that the operator $div$ maps  ${\cal
Z}^1(X)$ into  ${\cal Z}^0(X)$ and  ${\cal B}^1(X)$ into  ${\cal B}^0(X)$.
(The
first of these facts is geometrically obvious. Really  ${\cal Z}^1(X)$
consists
of $Q$-invariant   vector fields and the $S$-structure on $X$ is
$Q$-invariant.
Therefore $div$ maps  ${\cal Z}^1(X)$ into the space  ${\cal Z}^0(U)$
of
$Q$-invariant functions on $X$. Both facts can be checked by means
of direct
calculation.) We see that the operator $div$ acting from $Vect\  (X)$
into
$C(X)$ induces an operator acting from  ${\cal H}^1(X)$ into  ${\cal
H}^0(X)$;
let us denote this operator by $div_{{\cal H}}$.

   {\bf Lemma 4.} In the case $X=U$ the operator $div_{{\cal H}}$
coincides
with the operator $div:Vect\  (R_U)\rightarrow C(R_U)$ corresponding
to the
$S$-structure induced in $R_U$. (We use the identifications  ${\cal
H}^1(U)=Vect (R_U),\   {\cal H}^0(U)=C(R_U)$). In the general case
we
have a
commutative  diagram

 \[
  \begin{array}{ccc}
   {\cal H}^1(X)&
       \stackrel{div_{{\cal H}}}{\longrightarrow}&  {\cal H}^0(X)\\
       \Big\downarrow\vcenter{%
       \rlap{$\scriptstyle\rho_1$}}&
        &\Big\downarrow\vcenter{%
	\rlap{$\scriptstyle\rho_0$}}\\
	 Vect \,   (R)&\stackrel{div}{\longrightarrow}&C(R)
   \end{array}
   \]

   The first statement follows immediately from the consideration
above. The
second statement can be derived from the first one.

   {\bf Lemma 5.} If a function $f$ on a $QS$-manifold $X$ can be
represented
in the form $f=Q\varphi$ (i. e. $f\in {\cal B}^0(X)$), then the integral of
$f$
over $X$ vanishes. If $Y$ is empty the same is true for every
$Q$-invariant
function $f$ on $X$ (i. e. for every function $f\in {\cal Z}^0(X)$).

   The second statement of the Lemma was proved in [5]. It follows
immediately
from the fact that in the case at hand we can find a coordinate
system in
$X$ such that $Q=\partial /\partial \xi ^a$ where $\xi ^a$ is one of the
odd
coordinates. One can conclude from this statement that in the case
when $Y$ is
not empty the integral of $f$ over $X$ depends only on values of $f$
and its
derivatives at the points of $Y$; in other words this integral is
determined by
the function $f$ in an infinitesimal neighborhood of $Y$ (see[5]).

   The first statement can be proved easily in the case when the
function
$\varphi$ vanishes outside a compact set $K\subset U$ where $U$
satisfies the
conditions of Theorem 2 (i.e. $Q$ has the standard form (19) in $U$).
One can
reduce the general case to the simplest one using Theorem 2 and a
partition of
unity.

    \vskip .1in
   \centerline {\bf 5. $P$-manifolds and $SP$-manifolds.}
\vskip .1in

   Let is consider a (super)manifold $M$ with an odd closed 2-form
$\omega$. We
will say that $M$ is a  $P$-manifold (an odd symplectic manifold) if the
form
$\omega$ is non-degenerate. (The form $\omega={1\over
2}dz^i\omega_{ij}(z)dz^j$
determines an odd inner product $\omega(a,b)={1\over
2}a^i\omega_{ij}(z)b^j$ in
the tangent space $T_xM$ at every point $x\in M$. The form
$\omega$ is
non-degenerate if this inner product is  non-degenerate for all $x\in
M$.) One
can weaken the requirement of  non-degeneracy of $\omega$
assuming that for
every $x\in M$ the space $B_x=(T_xM)^{\perp}$ of vectors $b\in
T_xM$ satisfying
$\omega(a,b)=0$ for all $a\in T_xM$ is a superspace having
dimension
independent of $x$. In this case we will say that $M$ is a
pre-$P$-manifold (an
odd pre-symplectic manifold). One can prove that in a neighborhood of
every
point of  a pre-$P$-manifold $M$ there exists a local coordinate
system
$(\xi^a,\eta_a,\zeta^{\alpha} )$ such that $\omega=d\xi^ad\eta_a$.
(The
coordinates
$(\xi^a,\eta_a,\zeta^{\alpha} )$ are called Darboux  coordinates.) It
follows
immediately from this fact that subspaces $B_x\subset T_xM$
determine a
foliation of $M$; this foliation will be denoted by $\beta$. Let us say
that a
vector field $b$ on $M$ belongs to the set ${\cal B}$ if $b(x)\in B_x$
for
every $x\in M$. A function $F$ on $M$ is a constant on every leaf of
the
foliation $\beta$ iff it is ${\cal B}$-invariant (i.e. $\hat{b}F=0$ for every
$b\in{\cal B}$). If the foliation is a fibration one can identify the set of
${\cal B}$-invariant functions on $M$ with the set of functions on the
base
$M^{\prime}$ of this fibration. It is easy to check that the form
$\omega$
induces a non-degenerate closed 2-form $\tilde {\omega}$ on
$M^{\prime}$,
therefore $M^{\prime}$ can be considered as a $P$-manifold; we say
that this
$P$-manifold corresponds to the pre-$P$-manifold $M$. (The form
$\tilde
{\omega}$ is determined by the condition $\omega=\pi^*\tilde
{\omega}$ where
$\pi$ denotes the natural projection of $M$ onto $M^{\prime}$.)

  If the foliation $\beta$ is not a fibration we cannot construct a
$P$-manifold $M^{\prime}$ corresponding to the pre-$P$-manifold
$M$ but for
every point $x\in M$ we can find a neighborhood $U$ such that a
$P$-manifold
$U^{\prime}$ does exist.

   We say that $L$ is a Lagrangian submanifold of a pre-$P$-manifold
$M$ if $L$
is a maximal isotropic submanifold of $M$ transversal to $B_x$ at
every point
$x\in L$. (We say that a submanifold $L$ is isotropic if for every two
vectors
$a,b\in T_xL$ we have $\omega(a,b)=0$. The transversality condition
means that
$T_xL\cap B_x=\emptyset$ for all $x\in L$.) For every $x\in M$ the
space
$T_xM/B_x$ is provided with a non-degenerate inner product; it is
obvious that
$L$ is a Lagrangian submanifold of $M$ iff the natural map of $T_xL$
into
$T_xM/B_x$ is an isomorphism and its image is a maximal isotropic
subspace. If
the foliation $\beta$ is a fibration then the image $\pi(L)$ of $L$ under
the
natural projection $\pi:M\rightarrow M^{\prime}$ is a Lagrangian
submanifold of
$M^{\prime}$. (More precisely $\pi$ determines a Lagrangian
immersion of $L$
into $M^{\prime}$ because $\pi(L)$ can have self-intersections.)

    One can characterize Lagrangian submanifolds of
a $(n|n)$-dimensional
$P$-manifold $M$ as $(k,n-k)$-dimensional isotropic submanifolds of
$M$.

    It is easy to verify that $L$ is a Lagrangian submanifold of a
pre-$P$-manifold $M$ iff for every point $x\in L$ one can find a
basis
$(e^1,...,e^n,f_1,...,f_n,g^1,...g^l)$ of $T_xM$ such
that $e^1,...,e^n$ is a
basis
of $T_xL\subset T_xM$,
  \begin {equation}
\omega(e^i,e^j)=\omega(f_i,f_j)=\omega(e^i,g^{\alpha})=\omega(f_i,g^{
\alpha})=\omega(g^{\alpha},g^{\beta})=0,\ \  \omega(e^i,f_j)=\delta_j^i
\label {eq: (207)}
 \end {equation}
 (Here $1\leq i,\ j\leq n,\ \ 1\leq\alpha,\ \beta\leq l$.)

 If $M$ is a pre-$P$-manifold then the inner product in $T_xM$ permits
us to
lower indices. In other words, if we have a vector field $A^k$ we can
construct
a covector field $A_k=\omega_{kl}A^l$. If $M$ is a $P$-manifold the
correspondence between vector and covector fields is invertible: there
exists
an inverse matrix $\omega^{kl}$ for the matrix $\omega_{ab}$ and we
assign a
vector field $A^k=\omega^{kl}A_l$ to every covector field $A_l$. Hence
for
every function $H$ on $M$ we can construct a  vector field
$h^a=\omega^{ab}\partial_l H/\partial z^b$ (a Hamiltonian  vector field
with
the  Hamiltonian $H$). We get a map of the set $C(M)$ of functions on
$M$ into
the set $Vect(M)$ of  vector fields on $M$; this map will be denoted by
$K$.
Let us define the Poisson brackets $\{ G,H\}$ of two functions on $M$
by the
formula
 \begin {equation}
\{G,H\}={\partial_rG\over\partial
z^a}\omega^{ab}{\partial_lH\over\partial
z^b}.   \label {eq: (208)}
 \end {equation}
 Using the brackets (\ref {eq: (208)}) we can write the first order
differential operator $\hat {h}$ corresponding to a  vector field
$h=K(H)$ as
  \begin {equation}
 \hat {h}G=\{ G,H\}={\partial _rG\over \partial z^a}h^a=
 h^a{\partial _lG\over \partial z^a}   \label {eq: (209)} .
 \end {equation}
 The set $C(M)$ can be considered as a Lie superalgebra with respect
to the
operation (\ref {eq: (208)}) and $K$ is a (parity reversing)
homomorphism of
this superalgebra into the Lie superalgebra $Vect(M)$. More precisely,
brackets
(\ref {eq: (208)}) obey
  \begin {equation}
 \{ G,H\} =-(-1)^{(\varepsilon_G+1)(\varepsilon_H+1)}\{ H,G\}
 \end {equation}
  \begin {equation}
 (-1)^{(\varepsilon_F+1)(\varepsilon_H+1)}\{ F,\{ G,H\}\} +{\rm cyclic\
perm.}F,G,H=0 \label {eq: (211)}
 \end {equation}
(graded anticommutativity and Jacobi identity). The operator $K$
transforms
brackets into a graded (anti)commutator of vector fields.

   The transformations preserving an odd closed $2$-form $\omega$
are called
$P(\omega)$-transformations or simply $P$-transformations. As we
mentioned
above a pre-$P$-manifold $M$ can be covered with coordinate
systems
$(\xi^a_{(i)}, \eta_a^{(i)}, \zeta_{(i)}^{\alpha})$ in such a way that the
form
$\omega$ specifying the pre-$P$-structure on $M$ has the standard
form
   $$\omega_0=d\xi_{(i)}^ad\eta_a^{(i)}$$
   in these coordinates. Hence a pre-$P$-manifold can be pasted
together from
superdomains by means of $P(\omega_0)$-transformations. This fact
can be
considered an alternative definition of pre-$P$-manifold.

  A one-to-one map $F$ of a domain $U$ with coordinates
$(\xi^a,\eta_a,\zeta^{\alpha})$ into a domain $\tilde{U}$ with
coordinates
$(\tilde{\xi}^a,\tilde{\eta}_a,\tilde{\zeta}^{\alpha})$ will be called an
$SP$-transformation if it is a $P$-transformation (i.e.
$F^*\tilde\omega_0=\omega_0$, where $\omega_0=d\xi^ad\eta_a,\
\tilde{\omega}_0=d\tilde{\xi}^ad\tilde{\eta}_a$) and the determinant of
the
matrix
  \begin{equation}
 \left( \begin{array}{cc}
 \partial\tilde{\xi}^a /\partial\xi^b, & \partial\tilde{\eta}_a /\partial\xi^b
\\
 \partial\tilde{\xi}^a /\partial\eta_b, &
 \partial\tilde{\eta}_a /\partial\eta_b
 \end{array} \right)       \label{eq:(213)}
 \end{equation}
 is equal to 1. It is easy to check that the $P$-transformation
$F:U\rightarrow\tilde{U}$ induces a $P$-transformation
$F^{\prime}:U^{\prime}\rightarrow\tilde{U}^{\prime}$ of corresponding
$P$-manifolds; the unimodularity of the matrix (\ref {eq:(213)}) means
that the
Jacobian of $F^{\prime}$ is equal to $1$. We define a
pre-$SP$-manifold $M$ as
a manifold pasted together from domains $U_i$ with coordinates
$(\xi^a_{(i)},\eta^{(i)}_a,\zeta^{\alpha}_{(i)})$ by means of
$SP$-transformations.  A pre-$SP$-manifold $M$ will be called an
$SP$-manifold if
$M$ is a $P$-manifold. In other words an $SP$-manifold can be
pasted together
from domains with coordinates $(\xi^a_{(i)},\eta^{(i)}_a)$ by means of
$SP$-transformations (i.e. $P$-transformations with Jacobian equal to
$1$). If
there exists a $P$-manifold $M^{\prime}$ corresponding to the
pre-$SP$-manifold $M$ then a pre-$SP$-structure on $M$ determines
an
$SP$-structure on $M^{\prime}$ and conversely an $SP$-structure on
$M^{\prime}$
determines a  pre-$SP$-structure on $M$. There exists a natural
volume element
(an $S$-structure) in an $SP$-manifold. If a manifold $M$ is provided
with a
$P$-structure and an $S$-structure simultaneously we introduce an
operator
$\Delta ={1\over 2}div\cdot K$ on the space $C(M)$ as a composition
of
the
operators $K:C(M)\rightarrow Vect\ M$ and $div:Vect\ M\rightarrow
C(M)$. For an
$SP$-manifold we have $\Delta^2=0$. One can prove that conversely
in the case
when $\Delta^2=0$ a  manifold provided with a $P$-structure and an
$S$-structure is an  $SP$-manifold [2].

 If $M$ is a  pre-$SP$-manifold we can define the operator $\Delta$ as
an
operator acting on the space $C(M,\cal {B})$ of $\cal {B}$-invariant
functions
on $M$. If there exists an  $SP$-manifold $M^{\prime}$ corresponding
to $M$
this fact follows from the identification $C(M,{\cal {B}})=C(M^{\prime})$.
In
the general case we can use the fact that for every point $x\in M$
there
exists
a neighborhood such that $C(U,{\cal {B}})=C(U^{\prime})$.

  If $L$ is a Lagrangian  submanifold of an  pre-$SP$-manifold $M$
one can
introduce a volume element (an $S$-structure) in $L$ as follows. If
$e^1,...,e^n$ is a basis in the tangent space $T_xL,\  x\in L$, we
extend it to
a basis $e^1,...,e^n,f_1,...,f_n,g^1,...,g^l$ of $T_xM$ satisfying (\ref
{eq:
(207)}). Then the projections
$\pi_xe^1,...,\pi_xe^n,\pi_xf_1,...,\pi_xf_n$ of
the first $2n$ vectors of this basis constitute a basis in
$T_{\pi(x)}M^{\prime}$. (Recall that $\pi$ denotes the natural projection
of
$M$ onto the  $SP$-manifold $M^{\prime}$ corresponding to the
pre-$SP$-manifold $M$). We define the measure in $T_xL$ by the
formula
  \begin {equation}
 \lambda(e^1,...,e^n)=\mu(\pi_xe^1,...,\pi_xe^n,\pi_xf_1,...,\pi_xf_n)^{1/
2}
\label {eq: (215)}
 \end {equation}
where $\mu$ is the measure in $T_{\pi(x)}M^{\prime}$. (It is possible
that the
 $SP$-manifold $M^{\prime}$ does not exist. In this case we have to
modify the
definition above by replacing $M$ with a neighborhood $U$ of $x\in
L$ such that
$U^{\prime}$ does exist.)

  The formula (\ref {eq: (215)}) determines the volume element in $L$
up to a
sign; the choice of the sign is discussed in [2].

  One can prove the following statement.

  {\bf {Lemma 6.}} If $M$ is a pre-$SP$-manifold, $L$ is a Lagrangian
submanifold, $H\in C(M,{\cal B})$ and $\Delta H=0$ then
$$\int\limits _{L_1} Hd\lambda_1=\int\limits _{L_2} Hd\lambda_2$$
where $L_1$ and $L_2$ are Lagrangian submanifolds of $M$  such
that $m(L_1)$ is
homologous to $m(L_2)$ in $m(M)$.
  If there exists an $SP$-manifold $M^{\prime}$ corresponding to $M$
the Lemma
follows immediately  from the identification $C(M,{\cal
B})=C(M^{\prime})$ and
from the corresponding statement for $SP$-manifold (see[2]).

  If the function $H$ is represented in the form $H=\exp (-\hbar^{-1}S)$
the
equation $\Delta H=0$ is equivalent to the equation
  \begin {equation}
{1\over 2}\{ S,S\} =\hbar\Delta S       \label {eq: (216)}
 \end {equation}
 The equation (\ref {eq: (216)}) is called the quantum master equation
and
the
equation
 $$\{ S,S\} =0$$
 is known as the classical master equation or simply the master
equation. If
the
function $H=\exp (-\hbar^{-1}S)$ obeys $\Delta H=0$ for all $\hbar$
then
  \begin {equation}
 \{ S,S\} =0,\  \Delta S=0
 \end {equation}
  \vskip .1in
   \centerline {\bf 6. $QS$-manifolds, $QSP$-manifolds and the master
equation.}
\vskip .1in
 Let us suppose that a $P$-manifold $X$ is provided with an odd
vector field
$Q$ satisfying $\{ Q,Q\} =0$ (with a $Q$-structure). We will say that
$X$ is a
$QP$-manifold if the $P$-structure  is specified by an odd
non-degenerate
$Q$-invariant $2$-form $\omega$; this means that the Lie derivarive
$L_Q\omega$
of $\omega$ with respect to the vector field $Q$ vanishes:
$L_Q\omega =0$. In
terms of the odd Poisson bracket $\{ F,G\} $ on $X$ the
$Q$-invariance of
the $P$-structure means that
  \begin {equation}
 \hat {Q}\{ F,G\} =\{ QF,G\} +\{ F,\hat {Q}G\} (-1)^{\varepsilon_F+1}
\label
{eq: (217)}
 \end {equation}
 where $\hat {Q} $ stands for first order differential operator
$Q^a\partial
a$. Every solution $S$ to the master equation $\{ S,S\} =0$ on $X$
determines a
parity reversing first order differential operator $\hat {Q}=\hat {Q}_S$
by the
formula
  \begin {equation}
 \hat {Q}F=\{ F,S\} .      \label {eq: (218)}
 \end {equation}
 It satisfies $\hat {Q}^2=0$; therefore $S$ determines a $Q$-structure
on
$X$. It
is easy to check that (\ref {eq: (217)}) for the operator  (\ref {eq: (218)})
follows from the Jacobi identity (\ref {eq: (211)}).
 Hence every solution to the master equation determines a
$QP$-structure on
$X$. Conversely, at least locally every first order differential operator
$\hat{Q}$ satisfying  (\ref {eq: (217)}) can be represented in the form
(\ref
{eq: (218)}). (The assertion that an operator $\hat {Q}$ obeying (\ref
{eq:
(217)}) can be represented in the form (\ref {eq: (218)}) with
multivalued $S$
can be considered as a superanalog of the well known fact that a
vector
field
preserving the Poisson brackets on a symplectic manifold can be
obtained
from
a multivalued Hamiltonian.) We see that every $QP$-structure on a
$P$-manifold
$X$ corresponds to a multivalued solution of the master equation.

 To be more precise, if $\hat {Q}^2=0$ then for the corresponding $S$
we
have $\{
S,S\} =\gamma$ where $\gamma$ is an odd constant. However
$\gamma =0$ if $S$
has at least one critical point or if we don't have auxiliary odd
parameters.
(In mathematical language this means that $S$ is a superfunction on a
supermanifold. Physicists speaking about a supermanifold often have
in mind a
family of supermanifolds; i. e. they allow dependence of all functions
on
auxiliary parameters.)

  Let us suppose that a manifold $X$ is provided with an
$SP$-structure and
that $S$ is a solution of the master equation $\{ S,S\} =0$ satisfying
the
condition $\Delta S=0$. One can prove that in this case the operator
$\hat
{Q}=\hat {Q}_S$ anticommutes with $\Delta$. The proof is based on
the
formula [3]
  \begin {equation}
 \Delta (F\cdot G)=\Delta F\cdot G+(-1)^{\varepsilon (F)}F\cdot \Delta
G+(-1)^{\varepsilon (F)}
 \{ F,G\} .    \label {eq: (219)}
 \end {equation}
 Taking $F=S$ and applying $\Delta$ to (\ref {eq: (219)}) we obtain
 $$0=\Delta(S\cdot\Delta G)+\Delta\{ S,G\} .$$
 Applying (\ref {eq: (219)}) once more we get
  \begin {equation}
 0=\{ S,\Delta G\} +\Delta\{ S,G\} .     \label {eq: (220)}
 \end {equation}
 In other words,
  \begin {equation}
 \hat{Q}\Delta +\Delta\hat {Q}=0.      \label {eq: (221)}
 \end {equation}
 Conversely, if an odd first order differential operator $\hat {Q}$
anticommutes with $\Delta$ then it corresponds to a function $S$
satisfying the
master equation and the condition $\Delta S=0$. One can say that
such an
operator $Q$ preserves the $SP$-structure. Using this interpretation
we
can give a
little bit different proof of (\ref {eq: (221)}). Namely the condition $\Delta
S=0$ can be written in the form $div Q=0$ and therefore can be
interpreted as
$Q$-invariance of the $S$-structure on $X$. Taking into account that
$Q$-invariance
of the $P$-structure on $X$ follows from the master equation we see
that
$\Delta$
is $Q$-invariant i.e. $\{ Q,\Delta\} =0$.

   We will use the term $QSP$-manifold for an $SP$-manifold with a
$Q$-structure if the operators $\hat{Q}$ and $\Delta$ anticommute.

   Let us consider now the set $Y$ of all critical points of a function $S$
satisfying the master equation $\{ S,S\} =0$ on a $P$-manifold $X$. In
other
words we consider a $QP$-manifold $X$ and the set $Y=\{ x\in
X|Qx=0\}$. As in
every $Q$-manifold for every point $x\in Y$ we have an odd linear
operator
$Q_x$ in $T_xX$ satisfying $Q_x^2=0$. The $P$-structure on $X$
determines  an
inner product $\omega(a,b)={1\over 2}a^i\omega_{ij}b^j$ in $T_xX$; it
is easy
to check that the operator $Q_x$ is skew-symmetric with respect to
this
inner
product. Therefore the image $B_x$ of $Q_x$ is an orthogonal
complement to the
kernel $Z_x$ of $Q_x$ in this inner product. This means that the
restriction of
the inner product $\omega(a,b)$ to $Z_x$ is degenerate, but it induces
a
non-degenerate inner product in $H(T_x,Q_x)=Z_x/B_x$. We always
consider the
case when $Y$ is a manifold, $Z_x$ and $B_x$ are superspaces with
dimension
independent on $x\in Y$. In this case the restriction of the form
$\omega$ to
$Y$ determines the structure of an odd pre-symplectic manifold (a
pre-$P$-structure)
in $Y$. It follows from the facts mentioned above that the foliation of
$Y$
generated by the $Q$-structure on $X$ coincides with the foliation
induced by
the pre-$P$-structure on $Y$. If the foliation of $Y$ specified by the
subspaces
$B_x\subset Z_x=T_x(Y)$ is a fibration we get a $P$-structure on the
base
$Y^{\prime}$ of this fibration.

    Let us consider now an $SP$-manifold $X$ and a function $S$ on
$X$
satisfying $\{ S,S\} =0,\  \Delta S=0$. (Recall that these data determine
a
$QSP$-structure on $X$.) In this case the manifold $Y^{\prime}$ is
provided
simultaneously with a  $P$-structure because $X$ is a $QP$-manifold
and with a
$S$-structure because $X$ is a  $QS$-manifold.

    {\bf {Theorem 3.}} The  manifold $Y^{\prime}$ corresponding to a
$QSP$-manifold $X$ is an  $SP$-manifold (i.e. the $S$-structure and
$P$-structure on $Y^{\prime}$ determine an $SP$-structure on
$Y^{\prime}$.)

    To prove this theorem it is necessary to check that the operator
$\Delta={1\over 2}div \cdot K$ where $div:Vect(Y^{\prime})\rightarrow
C(Y^{\prime})$ is determined by the $S$-structure on $Y^{\prime}$ and
$K:C(Y^{\prime})\rightarrow Vect(Y^{\prime})$ is determined by the
$P$-structure on $Y^{\prime}$ obeys $\Delta^2=0$. It is sufficient to
check
this fact locally. This means that we can replace $X$ by $U$ where
$U$ is
chosen in such a way that the maps $\rho_0$ and $\rho_1$ determine
isomorphisms
${\cal H}^0(U)=C(U^{\prime}),\  {\cal H}^1(U)=Vect(U^{\prime})$.

It is easy to check that the map $K:C(U)\rightarrow Vect (U)$
determined by
the $P$-structure on $U$ transforms ${\cal Z}^0(U)$ into ${\cal
Z}^1(U),\
{\cal B}^0 (U)$ into ${\cal B}^1(U)$ and therefore generates a map
$K_{{\cal H}}$ from ${\cal H}^0(U)$ into ${\cal H}^1(U)$. One can
verify
that after the identifications ${\cal H}_0(U)=C(U^{\prime}),\
{\cal H}^1(U)=Vect (U^{\prime})$ the map $K_{{\cal H}}$ coincides with
the
map $K:C(U^{\prime})\rightarrow Vect(U^{\prime})$ determined by the
$P$-structure on $U^{\prime}$. Now one can apply Lemma 4 to prove
that
the operator
$\Delta_{{\cal H}}={1\over 2}div_{{\cal H}}\cdot K_{{\cal H}}$ coincides
with
$\Delta={1\over 2}div\cdot K$ after the identification ${\cal
H}^0(U)=C(U^{\prime})$.
It is evident that $\Delta^2_{{\cal H}}=0$, hence $\Delta^2=0$ and
$Y^{\prime}$ is an $SP$-manifold.

  The same arguments lead to a little bit more general result valid
also in the case when the foliation of $Y$ is not a fibration.

{\bf Theorem 3$^{\prime}$.} The manifold $Y$ corresponding to a
$QSP$-manifold
$X$
is a pre-$SP$-manifold.
\vskip .1in
{\bf 7. Torsion in linear $P$-manifolds.}
\vskip .1in
   Let suppose now that $E$ is provided with a non-degenerate odd
differential  form
$\omega={1\over 2}dz^i\omega_{ij}dz^j$ where $\omega _{ij}$ are
constants.
The
form $\omega$ is obviously closed, so we can say that $E$ is provided
with a $P$-structure. One can say that $E$ is a linear $P$-manifold
(This
means that the $P$-structure on $E$ is translationally invariant).
   A Lagrangian subspace $L$ of $E$ is by definition
a maximal isotropic subspace of $E$ (i.e. a maximal subspace where
$\omega$ vanishes: $\omega(z,\tilde{z})=0$ for $z\in L,\tilde{z}\in L$).
One can always find Darboux coordinates, i.e. a coordinate
system
$(x^1,...,x^n,\xi_1,...,\xi_n)$ in $E$ such that $\omega$  has the form
\begin {equation}
\omega=dx^i d\xi_i
\end {equation}
where $x^i$ and $\xi_i$ have opposite parity; moreover for every
Lagrangian
subspace $L$ one can find Darboux coordinates in such a way that
$L$ is
singled out by the equations $\xi_1=...=\xi_n=0$. The dimension of
Lagrangian subspace is equal to $(k,n-k)$ where $0\leq k\leq n$. If the
notations are chosen in such a way that $x^1,...,x^n$ are even,
$\xi_1,...\xi_n$ are odd, the
 Lagrangian subspace of dimension $(k,n-k)$ can be singled out in an
 appropriate coordinate system by equations $x^{k+1}=...=x^n=0,\
 \xi_1=...=\xi_k=0$. The difference between the even dimension and
the
 odd dimension of
a Lagrangian subspace  $L$ will be denoted by $d$ (i.e.
$d=k-(n-k)=2k-n$).

  One can apply the general theory of $P$-manifolds to the linear
$P$-manifold $E$. In particular, one can define Poisson brackets and
the Hamiltonian vector field $K_S$ corresponding to a function $S$.

  If $S$ is an even quadratic function,
the corresponding vector field
 $S={1\over 2}z^iS_{ij}z^j$
has the form
\begin {equation}
K^i(z)=\omega^{ij}S_{jk}z^k=\hat{S}_k^iz^k
\end {equation}
where
\begin {equation}
\hat{S}_k^i=\omega^{ij}S_{jk}
\end {equation}
can be considered as the matrix of the parity reversing linear operator
$\hat{S}$ acting on $E$. In more invariant terms we can say that
\begin {equation}
S(z)=\omega(z,\hat{S}z)  \label {eq:(23)}
\end {equation}
The bilinear form $S(z,w)$ corresponding to the quadratic form $S(z)$
can
be represented as
\begin {equation}
S(z,w)={1\over 2}z^iS_{ij}w^j=\omega(z,\hat{S}w).
\end {equation}
If $\{ S,S\}=0$ it follows from (29) that $\hat{S}^2=0$ and
therefore we can consider the homology group $H$ and the torsion.
(We always
suppose that $Z= Ker \hat {S}$ and $B= Im \hat {S}$ can be
considered as linear
(super) subspaces of $E$). One can say a quadratic function $S$
satisfying $\{
S,S\} =0$ in a linear $P$-space $E$ determines a linear
$QP$-structure
in $E$.

  Let us fix a volume element $\mu$ in $E$. If $L$ is a Lagrangian
subspace in $E$ then one can define a volume element in $L$ by the
formula
\begin {equation}
\lambda(e_1,...,e_n)=\mu(e_1,...,e_n,f^1,...,f^n)^{1/2}  \label {eq:(25)}
\end {equation}
where $e_1,...,e_n$ constitute a basis in $L$ and $f^1,...,f^n$ are
vectors in $E$ satisfying $\omega(e_i,f^j)=\delta_i^j$. In the case
when $L$ is singled out by equations $\xi_1=...=\xi_n=0$ in Darboux
coordinates and $\hat{e}_i=e_i^j\partial/\partial x^j$ we can take
$f^j=e_i^j\partial /\partial \xi_i$. The proof of
existence of vectors $f^1,...,f^n$ for a general Lagrangian subspace
can
be reduced to this simplest case. Moreover the same arguments show
that
one can impose an additional requirement $(f^i,f^j)=0$ on vectors
$f^i$;
then these vectors will be determined uniquely. Note that (\ref
{eq:(25)})
determines $\lambda$ only up to a sign and that (\ref {eq:(25)}) must
be modified by a constant factor $\pm i$ in the case when the number
part $m(\mu(e_1,...,e_n,f^1,...,f^n))$ of  $\mu(e_1,...,e_n,f^1,...,f^n)$
is negative. The ambiguity of sign will be neglected in all formulas
below;therefore we omit also other irrelevant sign factors.

{\bf Lemma 7.} Let $S$ denote a quadratic form of $E$ satisfying
$\{ S,S\}=0$ and $L$ denote a Lagrangian subspace of $E$ such that
the restriction of $S$ to $L$ is a non-degenerate quadratic form on
$L$.
Then the homology group constructed by means of the operator
$\hat{S}$
defined
by (\ref {eq:(23)}) is trivial and
\begin {equation}
Tor^{1\over 2}\hat{S}=(2\pi)^{d/2}\int_Le^{-S}d\lambda  \label {eq:(26)}
\end {equation}
where $\lambda$ denotes the volume element on $L$ corresponding
to a
volume element $\mu$ in $E$ and $Tor \hat{S}$ denotes the image of
$\mu$
by the isomorphism $\Lambda(E)=\Lambda(H)=R^{1,0}$. (Recall that
$d$ is
the difference between the even dimension and the odd dimension of
$L$.)

  To prove the Lemma let us take a basis $e_1,...,e_n$ satisfying
\begin {equation}
\lambda (e_1,...,e_n)=\mu(e_1,...,e_n,f^1,...,f^n)=1
\end {equation}
where $f^i$ obey $\omega(e_i,f^j)=\delta^j_i$ as above. The
coefficients
$a_{ij}$ in the decomposition
\begin {equation}
\hat{S}e_k=f^ja_{jk}+e_jb^j_k
\end {equation}
can be calculated by means of (43). Namely
\begin {equation}
a_{jk}=\omega(e_j, \hat{S}e_k)=S(e_j,e_k)
\end {equation}
We assume that the form $S$ is non-degenerate on $L$; therefore the
matrix
$a_{jk}$ is non-degenerate. We have some freedom in the choice of
$f^1,...,f^n$; namely we can replace $f^i$ by $f^i+k^i_je_j$ and
the condition
$\omega(e_j,f^i)=\delta^i_j $ will not be violated. Utilizing the
non-degeneracy of $a_{jk} $ one can check that this freedom is
sufficient
to get $b^j_i=0$. In such a way we obtain a basis
$e_1,...,e_n,f^1,...,f^n$
on $E$ obeying $\hat{S} e_k=f^ja_{jk}$. It follows immediately from
$\hat{S}^2=0$ and the non-degeneracy of $a_{jk}$ that $\hat{S}f^j=0$
for
this
basis. We obtain that the space $Z$ is spanned by $f^1,...,f^n$ and
coincides with $B$, and therefore $H=0$. One can apply (\ref {eq:(6)})
to
calculate the torsion; we get
\begin {equation}
Tor\hat{S}=\mu(e_1,...,e_n,\hat{S}e_1,...,\hat{S}e_n)=(\det a)^{-1}
\mu( e_1,...,e_n,f^1,...,f^n)=(\det a)^{-1}
\end {equation}
{}From the other side
\begin {equation}
\int_L e^{-S}d\lambda=(\det S)^{-{1\over 2}}(2\pi)^{-{d\over 2}}
\end {equation}
where $S_{ij}=S(e_i,e_j)$; we obtain (\ref {eq:(26)}) using (48).

One can generalize Lemma 1 to the case when the restriction $\tilde
{S}$
of $S$ is degenerate on $L$. Then the space $N$ of critical points of
$\tilde {S}$ is invariant with respect to  $\hat {S}$ (this follows
immediately from $\hat{S}^2=0$); we denote the operator  $\hat  {S}$
restricted to $N$ by $Q$. Let us consider the partition function
$Z_{\tilde
{S},Q}$ of
$\tilde {S}$ with respect to $Q$. In the case when the homology group
of $Q$ in $N$ is trivial one can prove that the homology group of
$\hat {S}$ in $E$ is trivial too and we can consider $Z_{\tilde {S},Q}$
and the
torsion
Tor  $\hat {S}$ as numbers. The generalization of (\ref {eq:(26)}) looks
as follows:
\begin {equation}
Z_{\tilde {S},Q}=(2\pi)^{d/2} Tor^{1/2}\hat{S}
\end {equation}
The same formula remains correct in some sense in the case when
$Q$ has
non-trivial homology in $N$. More precisely, one can prove

{\bf Lemma $7^{\prime}$.} Let $S$ denote a quadratic form on $E$
satisfying
$\{ S,S\} =0$. If $N$ stands for the space of critical points of the
restriction
$\tilde{S}$ of $S$ to a Lagrangian subspace $L$ and $Q$ denotes the
operator
$\hat{S}$ restricted to $N$ then there exists a canonical isomorphism
\begin {equation}
\Lambda(H(E,\hat{S}))=\Lambda(H(N,Q)\otimes
H(N,Q))=\Lambda(H(N,Q))^2
            \label {eq:(33)}
\end {equation}
(Here $H(E,\hat{S})$ stands for the homology group of the operator
$\hat{S}$
acting
in $E$ and $H(N,Q)$ for  the homology group of $Q$ acting on $N$.)
Using
the
identification of $\Lambda(H(E,\hat{S})$ and $\Lambda(H(N,Q))^2$ by
means
of (\ref {eq:(33)}) one can prove  that
\begin {equation}
(2\pi)^{-d}Tor \hat{S}=Z_{\tilde {S},Q}\otimes Z_{\tilde {S},Q}     \label
{eq:(34)}
\end {equation}
where $Z_{\tilde {S},Q}$ denotes the partition function of $\tilde{S}$
with
respect to $Q$
considered as an element of $\Lambda(H(N,Q))$.

To prove (\ref {eq:(34)}) we fix
the basis $e_1,...,e_n$ of $L$ in such a way that the first vectors
$e_1,...,e_k$ of this basis constitute a basis in $N$. The basis
$e_1,...,e_n$
of $L$ can be extended to a basis $e_1,...,e_n,f^1,...,f^n$ of $E$ by
means
of vectors $f^1,...,f^n$ satisfying $\omega(f^i,f^j)=0,\  \omega(e_j,f^i)=
\delta^i_j$.

  The operator $\hat{S}$  corresponding to the quadratic
functional $S$ has the form
\begin {equation}
\hat{S} e_i=f^j a_{ji}+e_j b_i^j
\end {equation}
\begin {equation}
\hat{S}f^i=f^j c^i_j+e_j d^{ji}
\end {equation}
It follows from our assumptions that $a_{ij}=a_{ji}=0$ for $1\leq i\leq k,\
1\leq j\leq n$. The subspace $N$ spanned by $e_1,...,e_k$ and the
subspace $M$
spanned by vectors $e_{k+1},...,e_n,\hat{S}e_{k+1},...,\hat{S}e_n$ are
invariant with respect to the operator $\hat{S}$. Therefore we can
consider the
coset space $F=E/(N+M)$ and the operator $\bar {S}$ induced by
$\hat{S}$ in
$F$. The vectors $\bar{f}^1,...,\bar{f}^k$ obtained from $f^1,...,f^k$ by
means
of natural projection $E\rightarrow F$ form a basis  in $F$; it is easy to
check that
\begin {equation}
\bar{S}\bar{f}^i=\bar{f}^jc^i_j.
\end {equation}
It is easy to check that the matrix $c$ can be obtained from the matrix
$b$ by
means of (super)transposition and parity reversion.
In invariant terms one can say that $F$ can be identified with $\Pi N^*$
and
$\bar{S}=\Pi Q^*$ (recall that the operator $\hat{S}$ acting on $N$ is
denoted
by $Q$). We see that
\begin {equation}
H(F,\bar{S})=\Pi H(N,Q)^*
\end {equation}
and therefore
\begin {equation}
\Lambda H(F,\bar{S})=\Lambda H(N,Q)
\end {equation}
It is evident that the subspace $M$ is acyclic with respect to $\hat{S}$
(this
follows from the non-degeneracy of $S$ on the subspace spanned by
$e_{k+1},...,e_n$) and therefore $H(M+N,\hat{S})=H(N,Q)$. Now we
can apply
the exact homology sequence of the pair $(E,M+N)$ to obtain that in
the
case
$H(N,Q)=0$ we have $H(E,\hat{S})=0$ and in the general case
\begin {equation}
\Lambda H(E,\hat{S})=\Lambda H(M+N,\hat{S})\otimes\Lambda
H(F,\bar{S})=\Lambda
H(N,Q)\otimes\Lambda H(N,Q).
\end {equation}
To prove (\ref {eq:(34)}) we have to apply (\ref {eq:(9)}). We obtain
\begin {equation}
Tor(E,\hat{S})=Tor(M+N,\hat{S})\otimes Tor(F,\bar{S})  \label {eq:(42)}
\end {equation}
where the torsions are calculated with respect to volume elements
determined by
the bases in $E,M+N$ and $F$ that we have chosen. It is easy to
check that
\begin {equation}
Tor(M+N,\hat{S})=Tor(M,\hat{S})\cdot Tor(N,Q),
\end {equation}
\begin {equation}
Tor(F,\bar{S})=Tor(N,Q)
\end {equation}
\begin {equation}
 Tor(M,\hat{S})=(\det \sigma)^{-1}
\end {equation}
where $\sigma$ denotes the matrix of the form $S$ on $M$ in the
basis
$e_{k+1},...,e_n$ (in other words $\sigma_{ij}=\omega(e_i,Se_j),\
k+1\leq i,\
j\leq n$). Combining these equations with (\ref {eq:(42)}) we get (\ref
{eq:(34)}).

 Let us apply Lemma $7^{\prime}$ to the important particular case
when the
space $N$ of critical points of $\tilde {S}$ consists of critical points of
$S$
(i. e. $N\subset Z= Ker\hat {S}$). It is easy to check that in this case
$Q=0$
and therefore $H(N,Q)=N$. We obtain

 {\bf Lemma $7^{\prime\prime}$.} Let us suppose that the quadratic
form $S$ on
$E$ satisfies $\{S,S\} =0$ and all critical points of the restriction $\tilde
{S}$ to a Lagrangian subspace $L\subset E$ are critical points of $S$
on $E$.
Then
 \begin {equation}
 \Lambda(H(E,\hat {S}))=\Lambda(N)\otimes \Lambda(N)         \label
{eq: (65)}
 \end {equation}
 and
 \begin {equation}
 (2\pi)^{-d} Tor \hat {S}=Z_{\tilde {S}}\otimes Z_{\tilde {S}} \label {eq:
(66)}
 \end {equation}
Here $Z_{\tilde{S}}$ denotes the partition function of $\tilde {S}$
considered
as an element of $\Lambda (N)$.

If $x^1,...,x^n,\xi_1,...,\xi_n$ are Darboux coordinates in $E$, then the
function
\begin {equation}
S=\xi_i\sigma ^{ij} \xi_j          \label {eq: (67)}
\end {equation}
obviously obeys $\{S,S\} =0$. One can prove that this form of $S$ is in
some
sense general.

{\bf Lemma 8.} If $S$ is a quadratic function on $E$ satisfying
$\{S,S\}=0$
then one can introduce Darboux coordinates
$x^1,...,x^n,\xi_1,...,\xi_n$
in $E$ in such a way that $S$ depends only on  $\xi_1,...,\xi_n$.

To give the proof we restrict to $Z=Ker \hat {S}$ the form $\omega$
specifying
the $P$-structure on $E$. We get a degenerate odd form  $\tilde
{\omega}$ on
$Z$; the space of null-vectors of $\tilde{\omega}$ coincides with $B=Im
\hat{S}$. Hence we can define a non-degenerate odd $2$-form
$\omega ^{\prime}$
on $H=Z/B$; the form $\omega ^{\prime}$ determines a linear
$P$-structure on $H$.
(Compare with the consideration of the more general case of an
arbitrary
$QP$-manifold in
Section 6.) Let us fix a Lagrangian linear subspace $\Lambda$ in $H$.
It is
easy to check that $L=\pi ^{-1}(\Lambda)$ where $\pi$ stands for the
natural
projection of $Z$ onto $H$ is a Lagrangian subspace of $E$. Really, it
is
evident that $L$ is isotropic. Using the relations $l_1=\lambda_1+b_1,\
l_2=\lambda_2+b_2,\  h=z_1-b_1=z_2-b_2,\  n=z_1+b_2=z_2+b_1$
we obtain
$l_1+l_2=n$; hence $L$ is a Lagrangian subspace. (Here we used the
notations
$dim E=(n,n),\  dim Z=(z_1,z_2),\  dim B=(b_1,b_2),\  dim H=(h,h),\
dim
\Lambda =(\lambda_1,\lambda_2),\  dim L=(l_1,l_2))$. Now we can
introduce \dar
\ $x^1,...,x^n,\xi_1,...,\xi_n$ in $E$ in such a way that $L$ is singled
out by
equations
 $\xi_1=...=\xi_n=0$. It is  easy to verify that $S$ depends only on
  $\xi_1,...,\xi_n$ in this coordinate system.

  One can reformulate Lemma 8 in a coordinate-free way.

  {\bf Lemma $8^{\prime}$.} If $S$ is a quadratic function on $E$
satisfying
$\{ S,S\} =0$ then one can find a Lagrangian subspace $M\subset E$
and a linear
projection of $E$ onto $M$ in such a way that $S(x)=S(\mu(x))$ and
the kernel
of $\mu$ is  a Lagrangian subspace of $E$.

  To deduce this statement from Lemma 8 it is sufficient to single out
$M$ by
equations     $x^1=...=x^n=0$ and define $\mu$ by the formula
  \begin {equation}
   \mu(x^1,...,x^n,\xi_1,...,\xi_n)=(0,...,0,\xi_1,...,\xi_n).
  \end {equation}
  It follows from our proof of Lemma 8 that $dim M=(m,n-m)$ where
$m$ is an
arbitrary number obeying $b_2\leq m\leq n-b_1$.

   Let us denote the set of critical points of the restriction of $S$ to $M$
by
${\cal K}$. It is easy to check that the set $Z$ of critical points of $S$
on
$E$ can be represented in the form $Z=\rho ^{-1}({\cal K})$, in other
words a
point $e\in E$ satisfies $\hat {S}e=0$ if and only if $\mu(e)\in {\cal K}$.

   Let us suppose that $\tilde {M}$ is a Lagrangian subspace of
$E$ such that
the projection $\mu$ considered as a map from $\tilde {M}$ into $M$ is
an
isomorphism. Then in the coordinate system $(x,\xi )$ where $\mu$ is
given by
(68) we can represent $\tilde {M}$ as a set of points
    $x^1,...,x^n,\xi_1,...,\xi_n$ where $\xi _i=\sigma _{ij}x^j,\  \sigma
_{ij}$ is a fixed matrix. It follows immediately from the description of the
set $Z$ that the critical points of $S$ considered as a function of $\tilde
{M}$ belong to $Y$. This observation permits us to prove:

    {\bf Lemma 9.} The critical points of $S$ restricted to generic
Lagrangian
subspace of dimension $(m,n-m)$ where $b_2\leq m\leq n-b_1$
belong to the set
$Z=Ker\  \hat {S}$.

    To give a proof of this Lemma we construct  a Lagrangian subspace
$M$ of
dimension $(m,n-m)$ and a map $\mu:\  E\rightarrow M$ by means of
Lemma
$8^{\prime}$. Then it remains to note that for a generic  Lagrangian
subspace of
the same dimension the projection $\mu$ of this subspace into $M$ is
an
isomorphism.

I am indebted to A. Givental, M. Kontsevich and E. Witten for useful
discussions.
  \vskip .1in
  \centerline{{\bf References}}
  \vskip .1in
  1. Batalin, I., Vilkovisky, G.: Gauge algebra and quantization. Physics
Letters, 102B, 27(1981);
  Quantization of gauge theories with linearly dependent generators.
Phys. Rev.
D29, 2567(1983)

  2. Schwarz, A.: Geometry of Batalin-Vilkovisky quantization.
Commun. Math.
Phys. (in press)

  3. Witten, E.: A note on the antibracket formalism. Preprint
IASSHS-HEP-9019

  4. Schwarz, A.: The partition function of a degenerate functional.
Commun.
Math. Phys. 67,1 (1979)

  5. Witten, E.: The $N$ matrix model and gauged WZW models.
Preprint
IASSNS-HEP-91126

  6.Dold, A.: Lectures on algebraic topology,1972,Springer
  \end {document}